# Towards Innovative Physical Layer Design for Mobile WSN Platforms

Malay Ranjan Tripathy and Priya Ranjan

Amity School of Engineering and Technology, Amity University Uttar Pradesh, Noida, India.

mrtripathy@amity.edu

**Abstract**

We today live in the era of dynamic and mobile wireless enabled platforms. This kind of stringent communication capability in the face of volatile and turbulent mobility demands a fresh look at physical layer in general and antenna design in particular. The dimension of the antenna is 30×35×1.6 $mm^3$. Multiplicity of bands is very useful for compatibility purposes where legacy robotic platforms generally operate in MHz range while latest robotic platforms are capable to handle GHz communication regimes and can pump data very at much greater speeds. Seven frequency bands are obtained at 700 MHz, 2.4 GHz, 3.6 GHz, 4.37 GHz, 5.8 GHz, 6.93 GHz and 7.7 GHz with bandwidth of 1.1 GHz, 0.7 GHz, 0.8 GHz, 0.23 GHz, 0.90 GHz, 0.19 GHz and 0.12 GHz respectively.

## 1. Introduction

Wireless sensor networks (WSN) are one of the most incredible innovations of our times. When fitted with mobile platforms, they can do wonders in various disciplines. Some of the leading applications in Indian context are measuring the quality of pollution in water bodies, reporting air quality as a function of time and space, various defense applications like remotely control survey and sensing vehicles in swarming or isolated fashion. This kind of stringent requirement demands a very special kind of rethink on physical layer design targeted towards IoT and other operational regimes.

Design of a miniaturized UHF-band Zigbee antenna applicable to the M2M/IoT has been reported in [1] while energy harvesting with directional dipole antenna in bandwidth of interest has been presented in [2]. Antenna for wearables has been developed in [3] and environmental monitoring gadgets equipped with WSN capabilities are demonstrated in [4]. Various other Antenna and their related power control mechanism along with supporting electronics has been reported in various papers [5-12].

Our motivation for this particular design was to demonstrate a return loss of -39.98 at 0.7 GHz with a bandwidth of 1.1 GHz and VSWR 1.02. We believe that such designs and related supporting electronics has the capability to change the way we communicate at physical layer. In particular, our return loss is also significant IoT frequency 2.4 GHz. We believe that this is another exciting design which will open new vista in this domain.

This work is organized as follows: Section II has the description of antenna design. Section III contains details of performance metrics and discussion of proposed antenna and section IV presents the conclusive remarks.

## 2. Antenna Structure and Design

The antenna is designed on FR4 substrate with a thickness of 1.6 mm, $\varepsilon_r$ = 4.4 and loss tangent 0.02. The top view of the antenna is shown in Fig. 1. The dimension of the antenna is 30×35×1.6 $mm^3$. Microstrip line feeding is used in this antenna. It has a structure which is a combination of split ring resonator and comb shaped antenna. This design is simple, compact and miniaturized. It seems promising to integrate with rest of the wireless sensor network board. The antenna is simulated by using HFSS 14 (High Frequency Simulation Software) by ANSOFT.

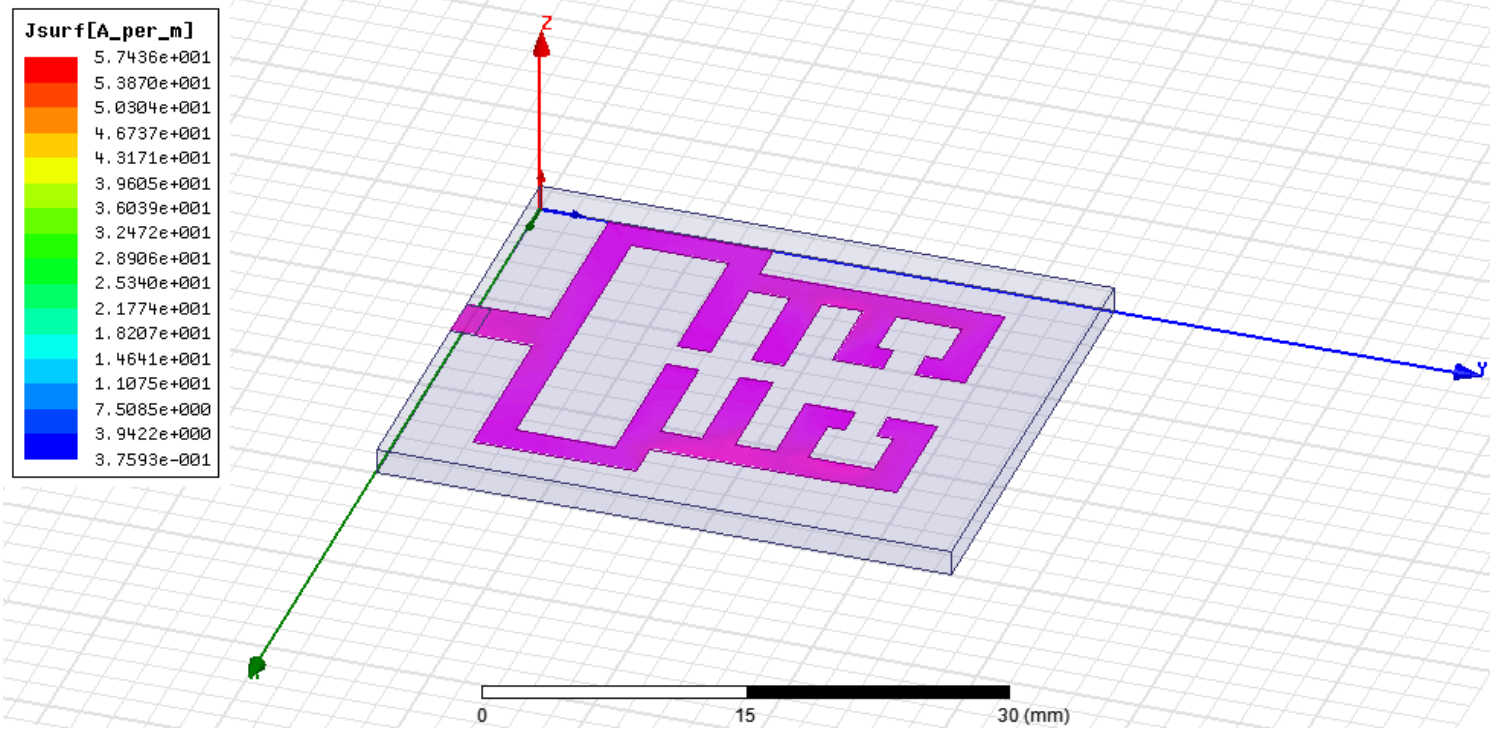

Fig. 1 Geometry of Proposed Antenna

## 3. Results and Discussion

The Return loss Vs. Frequency plot for the proposed antenna is shown in Fig.2. Seven interesting bands are obtained in this design. The band at 700 MHz is quite broad with impedance bandwidth of 1.1 GHz. It has return loss of -39.98 dB. This band is suitable for white space (470 – 698 MHz), GSM and underground communications. The band at 2.4 GHz is very important and it covers IOT, WLAN and ISM applications. The impedance bandwidth of this band is 0.7 GHz and return loss of – 28.35 dB. Third band at 3.6 GHz has good impedance bandwidth of 0.8 GHz and return loss -31.12 dB. This covers WiMAX applications. Fourth band at 4.37 GHz has the impedance bandwidth 0.23 GHz and return loss of -13.24 dB. Fifth band at 5.8 GHz has the impedance bandwidth 0.9 GHz with the return loss of -24.70 dB. The sixth band at 6.93GHz has impedance bandwidth of 0.19 GHz and return loss of -10.76 dB. The last seventh band is seen to have impedance bandwidth of 0.12 GHz and return loss of -10.36.The details of these frequency bands are mentioned in Table 1.

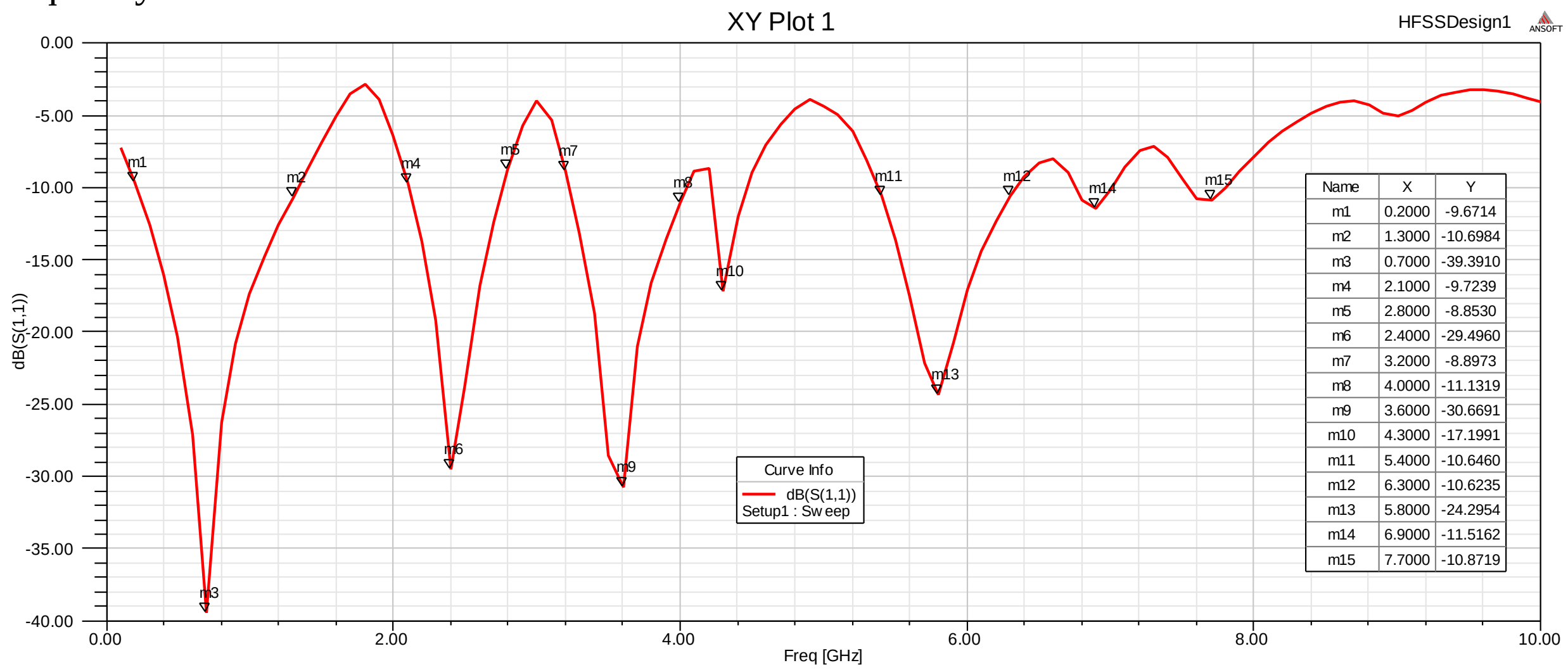

Fig. 2: S11 parameter of the antenna

Table 1 describes the return loss, impedance bandwidth and VSWR of proposed multiband antenna. Seven interesting bands are available with this antenna. The maximum impedance bandwidth of 1.1 GHz is seen at 0.7 GHz frequency with return loss of -39.98 dB, which is directly relevant for white space, GSM and underground communications.

Table 1: Antenna parameters at different frequency bands

| Sl. No. | Frequency (GHz) | Return Loss (dB) | Bandwidth (GHz) | VSWR |
|---|---|---|---|---|
| 1 | 0.7 | -39.98 | 1.1 | 1.02 |
| 2 | 2.4 | -28.35 | 0.7 | 1.08 |
| 3 | 3.6 | -31.12 | 0.8 | 1.08 |
| 4 | 4.37 | -13.24 | 0.23 | 1.6 |
| 5 | 5.8 | -24.70 | 0.90 | 1.12 |
| 6 | 6.93 | -10.36 | 0.19 | 1.80 |
| 7 | 7.7 | -10.36 | 0.12 | 1.96 |

Fig. 3 shows the VSWR Vs Frequency plot for proposed antenna. The bands at 0.7 GHz, 2.4 GHz, 3.6 GHz, 4.37 GHz, 5.8 GHz, 6.93 GHz and 7.7 GHz display VSWR 1.02, 1.08, 1.08, 1.6, 1.12, 1.8 and 1.96 respectively. This is also mentioned in the Table 1.

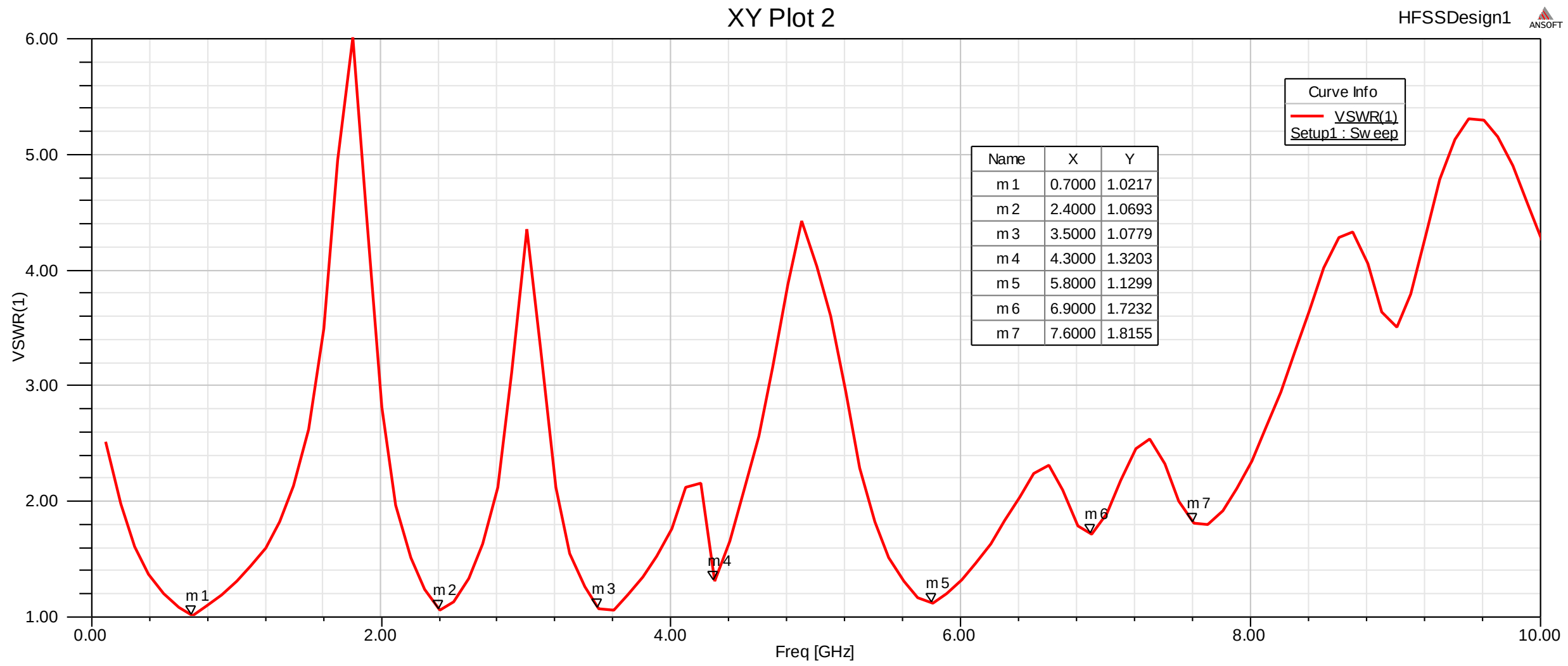

Fig. 3 VSWR Vs. Frequency plot of proposed Antenna

The simulated E plane radiation pattern of proposed multiband antenna at 700 MHz is shown in Fig. 4(a). The pattern for Φ = 0 is seen as directional but the pattern at Φ = 90 is fully bi-directional. Fig. 4(b) illustrates H plane radiation pattern at 700 MHz. The radiation pattern at θ = 0 is omnidirectional and the gain is less than the E plane pattern. The radiation pattern at θ = 90 is bi-directional with improved gain.

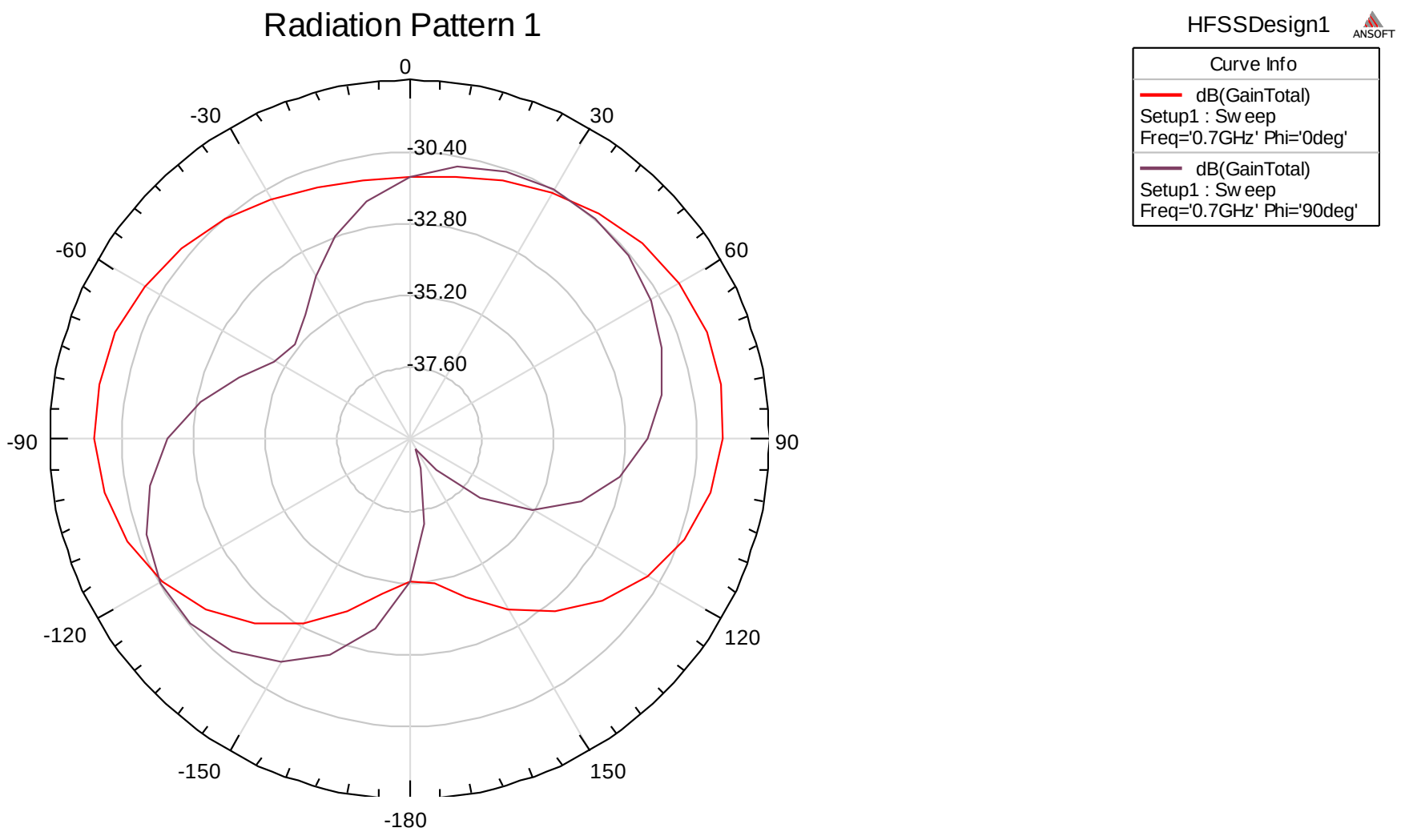

Fig. 4(a) E Field Radiation Pattern at 700 MHz

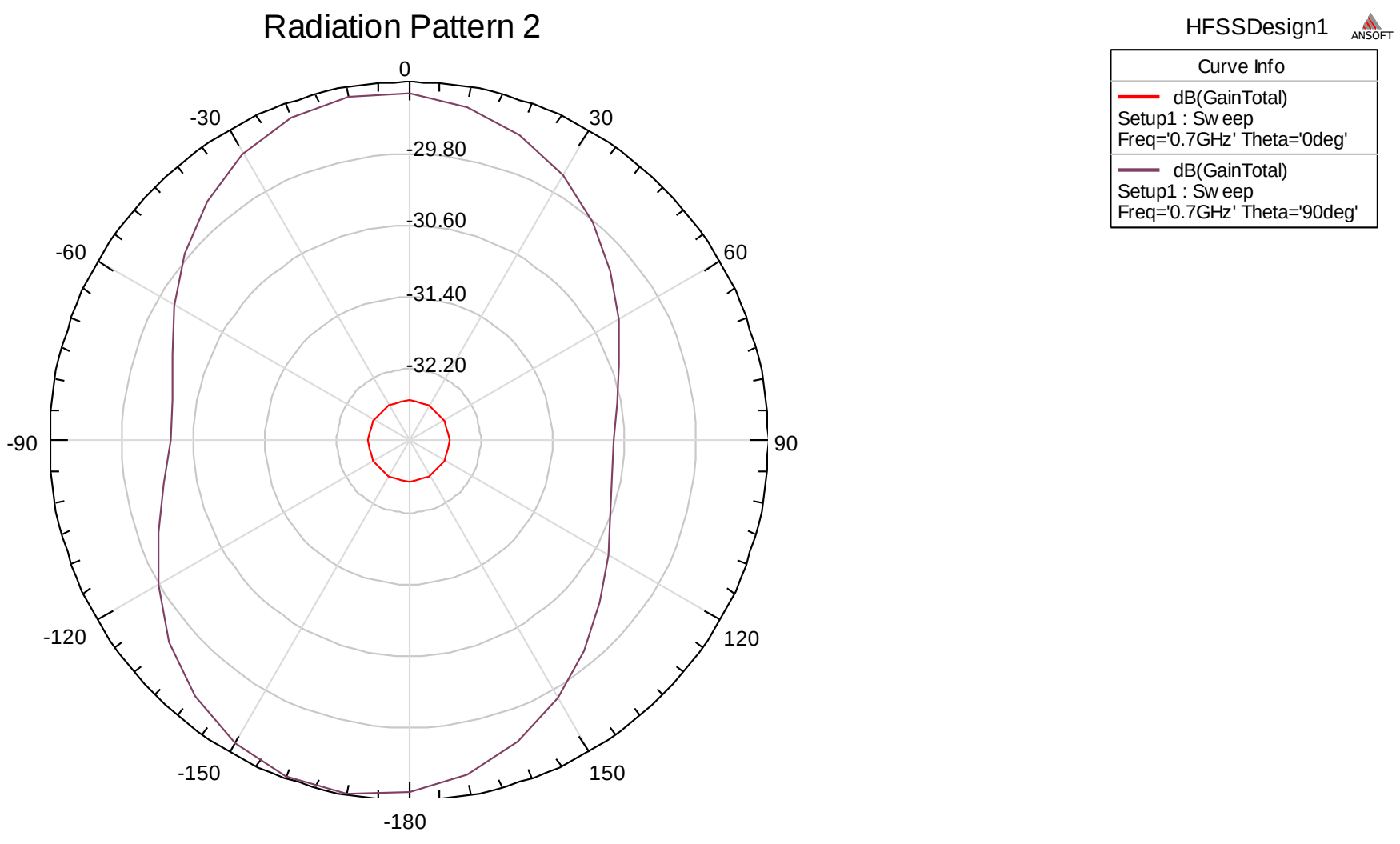

Fig. 4(b) H Field Radiation Pattern at 700 MHz

In Fig. 5(a) and 5(b), the simulated E & H plane radiation patterns of proposed multiband antenna at 2.4 GHz are shown respectively. The patterns for $\Phi = 0$ is seen to be nearly omni-directional but, for $\Phi = 90$ it is seen to be directional. The gain in both the cases are better than the Fig 4(a) and (b). In Fig. 5(b) it is seen that the radiation pattern for $\theta = 0$ is Omni-directional and for $\theta = 90$ is directional. But the relative gain is better in comparison to corresponding pattern at 700 MHz.

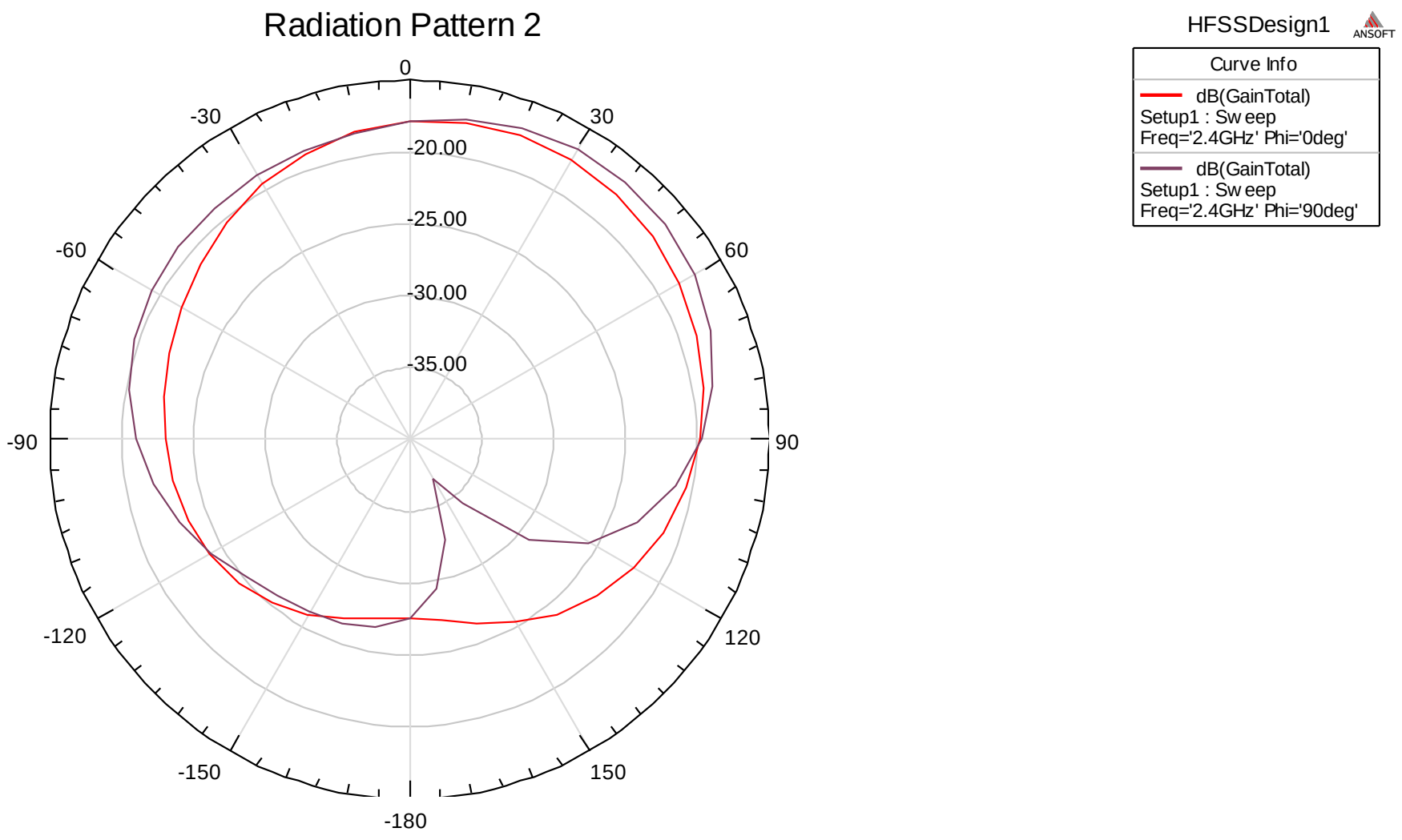

Fig. 5(a) E Field Radiation Pattern at 2.4 GHz

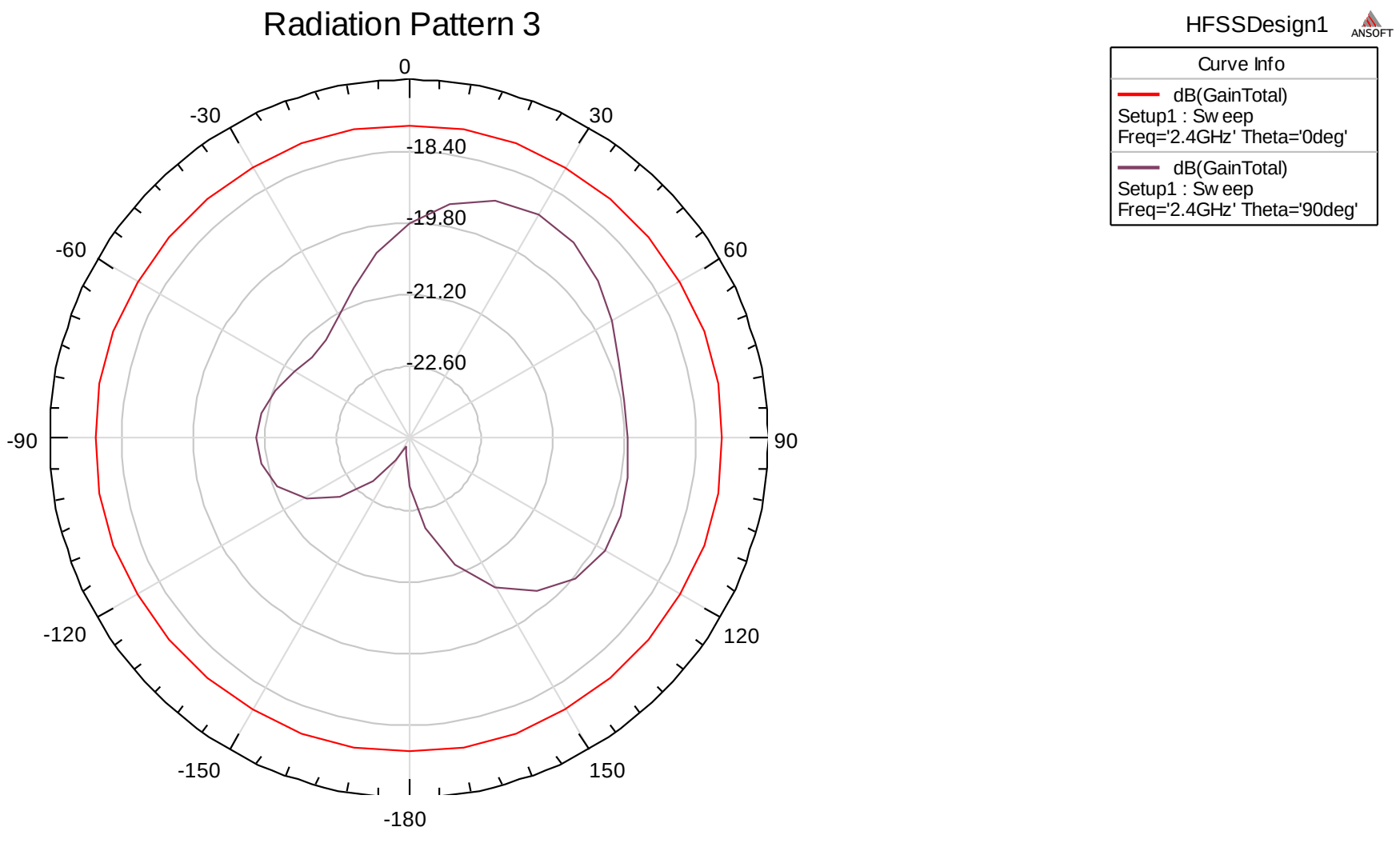

Fig. 5(b) H Field Radiation Pattern at 2.4 GHz

Fig. 6(a) and 6(b) show the simulated E & H plane radiation patterns of proposed multiband antenna at 3.5 GHz respectively. The patterns for Φ = 0 and Φ = 90 are seen to be similar and bi-directional. In Fig. 5(b) it is seen that the radiation pattern for θ = 0 is omni-directional with much less gain whereas, for θ = 90 the radiation pattern is directional with better gain

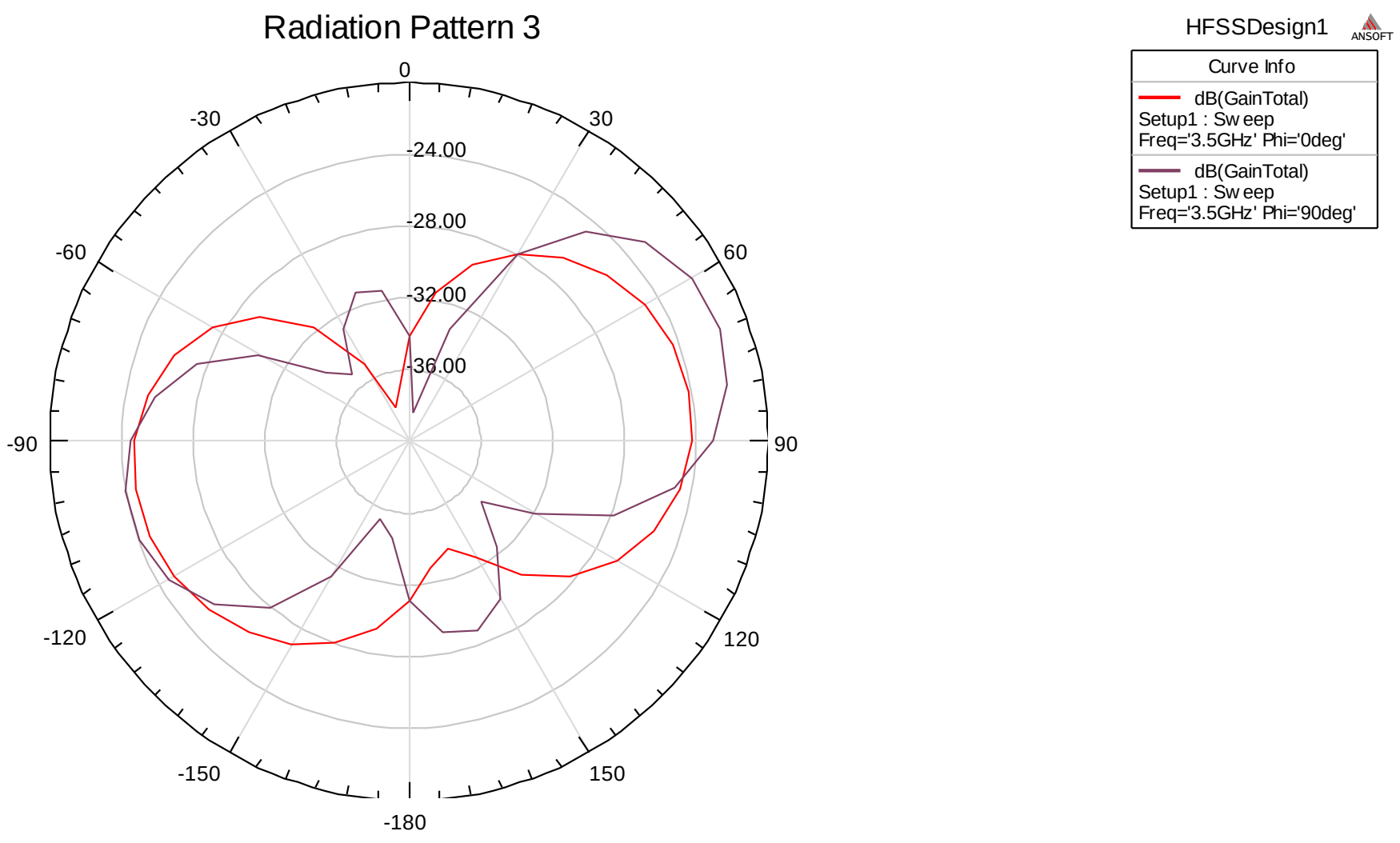

Fig. 6(a) E Field Radiation Pattern at 3.5 GHz

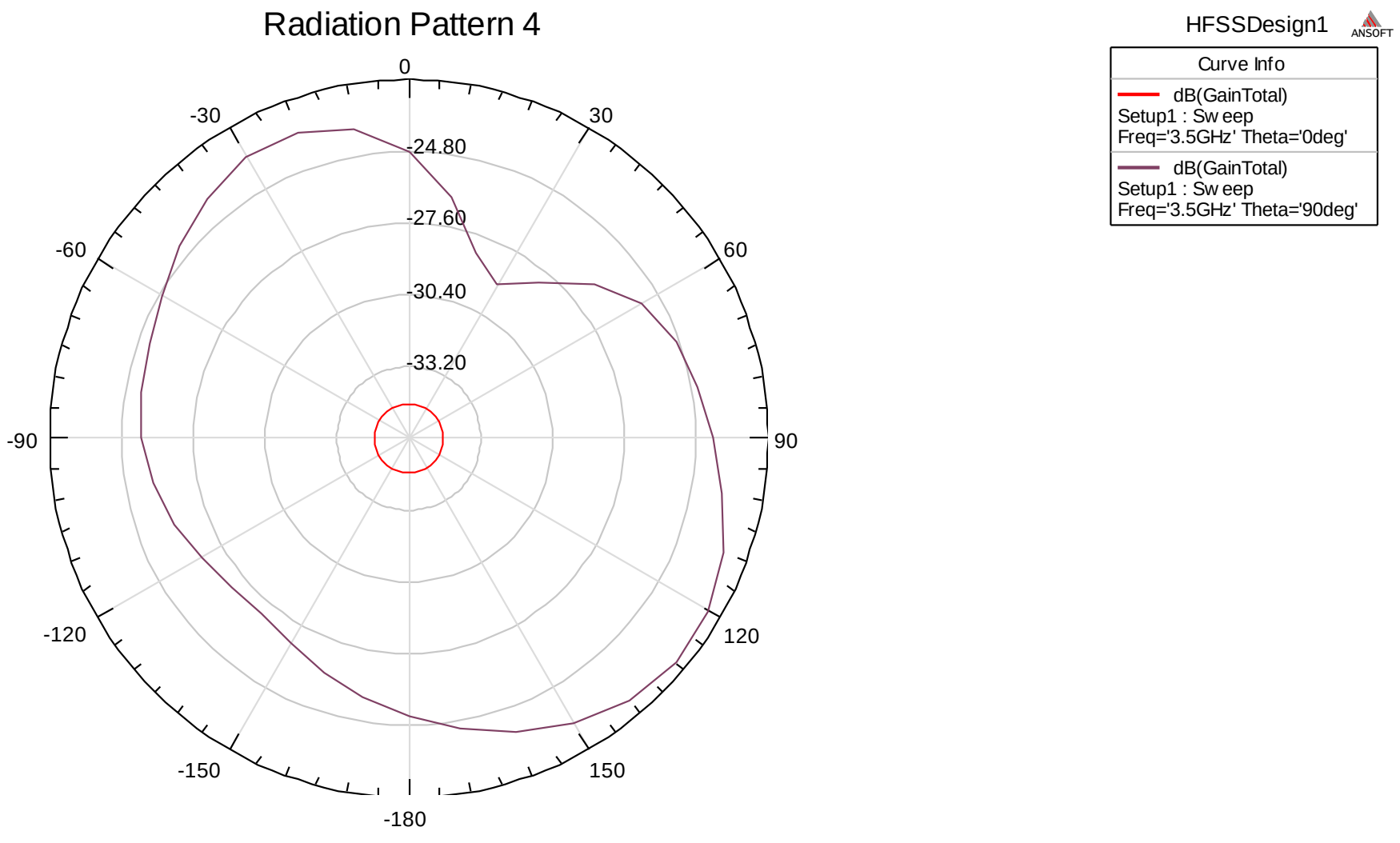

Fig. 6(b)  H Field Radiation Pattern at 3.5 GHz

The simulated E & H plane radiation patterns of proposed multiband antenna at 5.8 GHz are shown in Fig. 7(a) & 7(b) respectively. The patterns for Φ = 0 and Φ = 90 are seen to be directional. But the gains in both the cases are better than the patterns are shown at 700 MHz, 2.4 GHz and 3.5 GHz. In case of Fig. 7(b) radiation pattern for θ = 0 is Omni-directional and for θ = 90 is bi-directional. But the relative gain is better in comparison to rest of patterns shown in this paper.

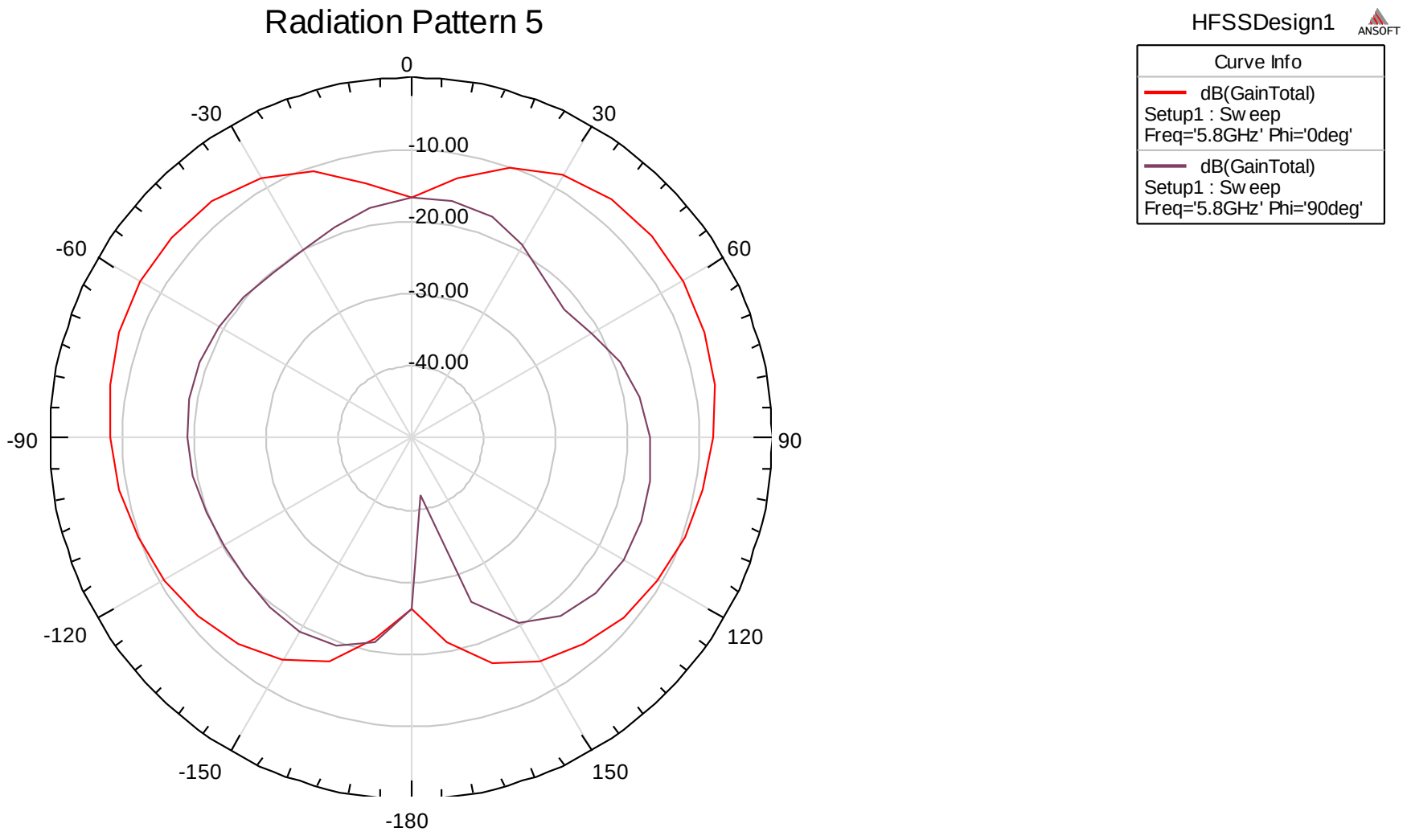

Fig. 7(a)  E Field Radiation Pattern at 5.8 GHz

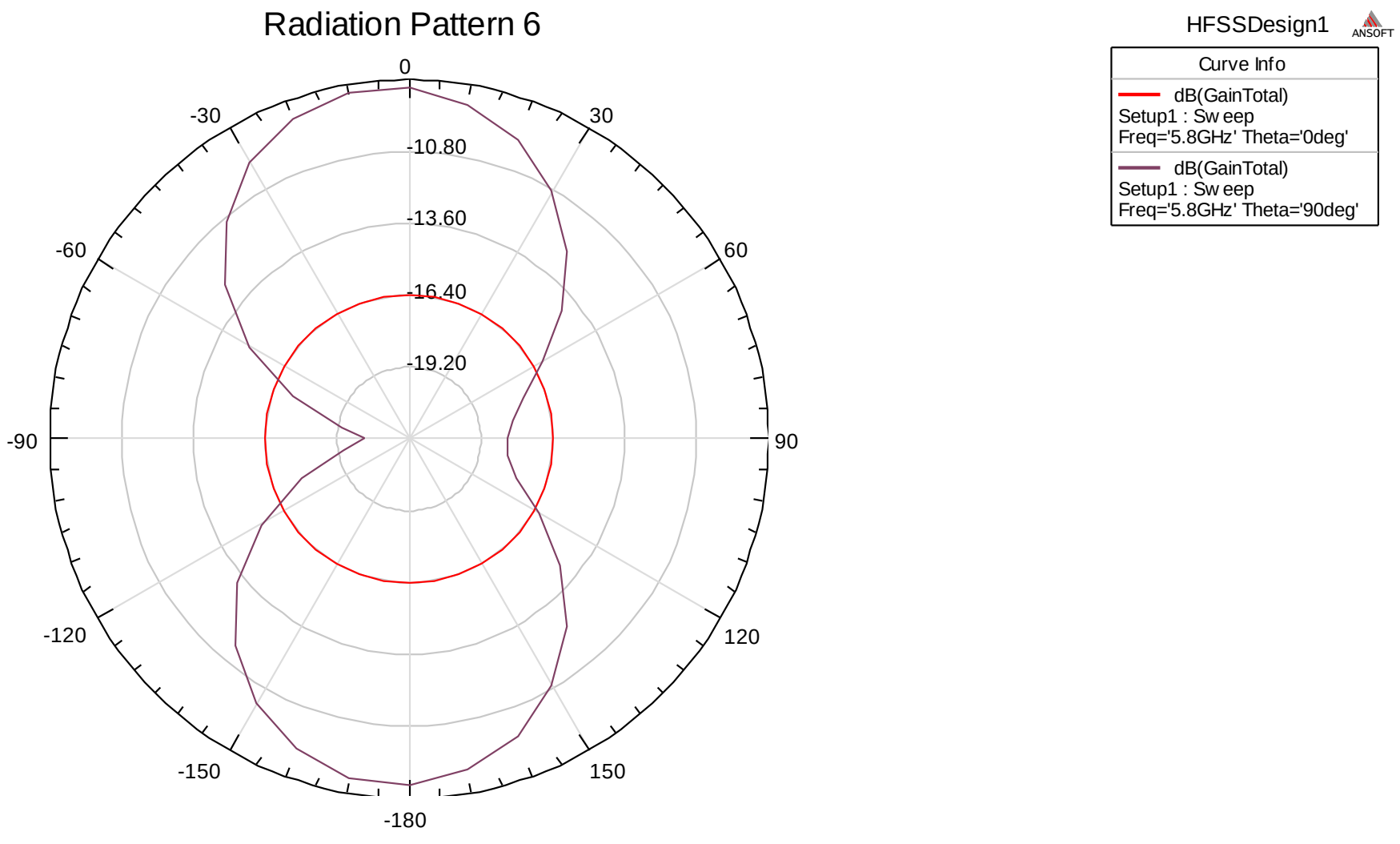

Fig. 7(b) H Field Radiation Pattern at 5.8 GHz

Fig. 8 shows the gain Vs. frequency plot of the proposed antenna. Though the gain at 700 MHz, 2.4 GHz, 3.5 GHz and 5.8 GHz remained –ve but the peak gain is observed as 4.14 dB at 9.1 GHz for $\Phi$ = 90, $\theta$ = 350.

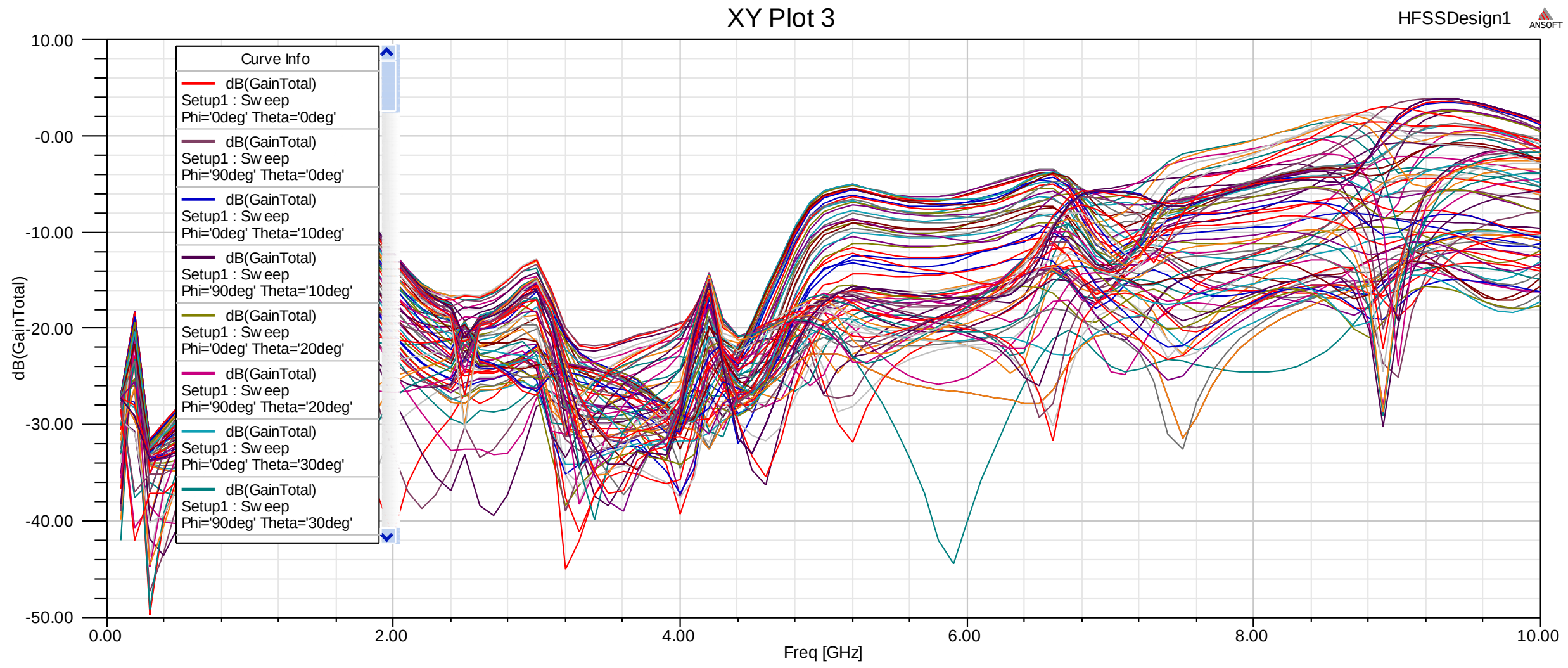

Fig. 8 Gain Vs Frequency Plots

The surface current density distribution on radiating patch at 2.4 GHz is shown in Fig. 9. It is seen that current density is distributed throughout the radiating patch. The coupling between feedline and rest of the patch is very strong. However, the current distribution is stronger at lower part of the split ring resonator region of the patch. The gain and other parameters can be improved by adjusting design parameters of antenna and ground plane of the proposed compact, simple and miniaturized multiband antenna.

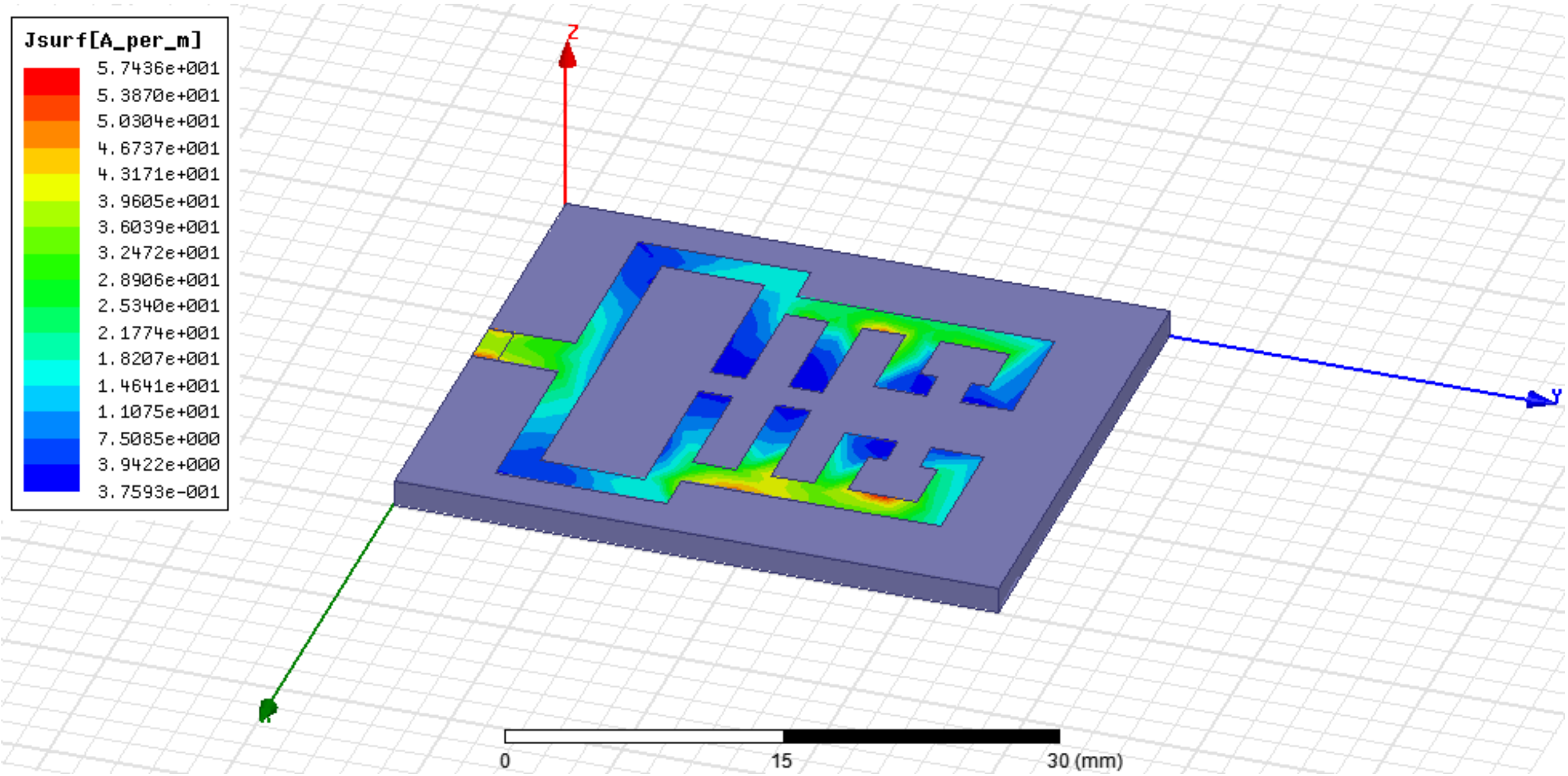

Fig. 9 Current Density distribution at 2.4 GHz

## 4. Conclusions

This paper has addressed the issues of designing simple, compact, miniaturized and multiband antenna for swarming robotic applications and their integration for multi-band operation. It bridges the much needed gap for multi-band antennas to support communication for robotic platforms. The possible integration with the robotic platform being designed is also discussed. The antenna is designed on FR4 substrate with a thickness of 1.6 mm, $\varepsilon_r = 4.4$ and loss tangent 0.02. The dimension of the antenna is $30 \times 35 \times 1.6$ mm$^3$. Multiplicity of bands is very useful for compatibility purposes where legacy robotic platforms generally operate in MHz range while latest robotic platforms are capable to handle GHz communication regimes and can pump data very at much greater speeds. Seven frequency bands are obtained at 700 MHz, 2.4 GHz, 3.6 GHz, 4.37 GHz, 5.8 GHz, 6.93 GHz and 7.7 GHz with bandwidth of 1.1 GHz, 0.7 GHz, 0.8 GHz, 0.23 GHz, 0.90 GHz, 0.19 GHz and 0.12 GHz respectively. Microstrip line feeding is used in this antenna. It has a structure which is a combination of split ring resonator and Comb shaped patch antenna. Future work includes real life demonstration with this antenna integrated with state of art robotic/vehicle platforms and next generation antenna design from lessons learnt from these experiments.